# Position Uncertainty Measures on the Sphere

D.A. Trifonov

Institute for nuclear research,
72 Tzarigradsko chaussee, 1784 Sofia, Bulgaria

November 21, 2018

**Abstract**

Position uncertainty (delocalization) measures for a particle on the sphere are proposed and illustrated on several examples of states. The new measures are constructed using suitably the standard multiplication angle operator variances. They are shown to depend solely on the state of the particle and to obey uncertainty relations of the Schrödinger–Robertson type. A set of Hermitian operators with continuous spectrum is pointed out the variances of which are complementary to the longitudinal angle uncertainty measure.

## 1. Introduction

Recently an interest is shown in the literature to the problem of a quantum particle on the circle [3, 4, 5, 7, 8, 9, 13] and on the sphere [4, 6, 10]. In [5, 6, 8, 10] overcomplete families of states (coherent states) for these systems are constructed.

One of the difficulties for these systems is the position (and momentum) uncertainty measures for the particle (or equivalently, the wave function spread measure). This is a consequence of the issue related the choice of the operator for the azimuthal angle $\varphi$. For a particle on the sphere there is a second problem, related to the non-hermiticity of the operator $-\mathrm{i}\partial/\partial\vartheta$, where $\vartheta$ is the longitudinal angle. The problem of correct definitions of uncertainty measures is (and should be) closely related to the construction and justification of the uncertainty relations (UR's), and of coherent and squeezed states as well.

From the Dirac correspondence rule between Poisson bracket $\{f,g\}$ of two classical quantities $f$ and $g$ and the commutator of the corresponding operators $\hat{f}$ and $\hat{g}$,

$$\{f,g\} \longrightarrow \mathrm{i}[\hat{f},\hat{g}] \tag{1}$$

it follows that $[\hat{p}_\varphi,\hat{\varphi}] = -\mathrm{i}$, where $\hat{\varphi}$ is the azimuthal angle operator, and $\hat{p}_\varphi$ is the angular momentum operator. Formally this commutation relation is satisfied by $\hat{\varphi} = \varphi$, and $\hat{p}_\varphi = -\mathrm{i}\partial/\partial\varphi$. Then one can write the standard Heisenberg-Robertson UR $(\Delta p_\varphi)^2(\Delta\varphi)^2 \geq 1/4$.

However on the eigenstates $\psi_m(\varphi)$ of $\hat{p}_\varphi$,

$$\psi_m(\varphi) = \exp(\mathrm{i}m\varphi)/\sqrt{2\pi}, \qquad m = 0, \pm 1, \ldots \tag{2}$$

the above UR breaks down.

The reasons for this contradiction is the fact, that $\hat{p}_\varphi$ is not Hermitian on the $\varphi$-transformed state $\psi'(\varphi) = \varphi\psi(\varphi)$, since $\psi'(\varphi)$ is no more invariant under translations on $2\pi$ (no more $2\pi$ invariant). Therefore many authors try to adopt some $2\pi$-invariant position operator [1, 3, 5, 9], or even another definition of the uncertainty on the circle [9]. However almost all of the associated uncertainty measures are not in good consistency with the localization of the particle on the circle [13].

In the case of a particle on the sphere the situation is even worth since in addition to the problems with the $\varphi$ and $p_\varphi$, one encounters the subtle of non-hermiticity of the operator $-i\partial/\partial\vartheta$, related to the longitudinal angle coordinate $\vartheta$, $0 \leq \vartheta \leq \pi$.

In this paper we provide an approach to these issues with minimal (in our opinion) deviation from the standard commutation relation and standard measure of uncertainty. In the case of a particle on the circle the main idea has been sketched in the second paper in [13], and developed in greater detail (providing some proofs and examples) in the first paper in [13].

In section 2 a brief review of the properties of main previous position uncertainty measures on the circle (based on $2\pi$-periodic operator $\hat{\varphi}$) is provided. In section 3 two different position uncertainty measures for a particle on the sphere are constructed and discussed. At $\vartheta = \pi/2$ the corresponding states and measures on a circle are recovered. The new measures are constructed using suitably the standard expressions of the first and second moments of the angle variable, calculated by integration over $2\pi$ intervals. They are of the form of positive state functionals, the values of which depend solely on the state considered. In section uncertainty relations (UR's) (of the Robertson–Schrödinger type) are established for the position delocalization measures and the appropriate *complementary measures*. The latters are of the form of standard variances of $\hat{p}_\varphi$ and new Hermitian operators $\hat{p}_{n\vartheta}$, $n = 1, 2, \ldots$.

## 2. Uncertainty measures on the circle

For a particle on the real line the standard measure of the position uncertainty is given by the second moment $(\Delta x)^2 := \langle (x - \langle x \rangle)^2 \rangle$ of the position operator $\hat{x} = x$, or equivalently by the standard deviation $\Delta x$. Mathematically both $\Delta x$ and $\langle x \rangle$ are one-to-one functionals on state space. The quantity $(\Delta x)^2$ is also called variance, or dispersion, of $x$ and is also denoted as $Dx$ or $M^{(2)}x$. The variance of $x$ is regarded as a measure of spread, or delocalization, of the state wave function $\psi(x)$. More precisely this is a measure of spread of the probability distribution $p(x) := |\psi(x)|^2$. Here the means $\langle x \rangle$ and $\langle x^2 \rangle$ are calculated by integration with respect to $x$: $\langle x \rangle = \int x|\psi(x)|^2 dx$.

However in the case of angle operator $\hat{\varphi} = \varphi$ it was not clear how to calculate and interpret the analogous quantity $\Delta\varphi$, since the operator $\hat{\varphi} = \varphi$ is not invariant under translation $\varphi \to \varphi + 2\pi$ (not $2\pi$-periodic), while the wave functions $\psi(\varphi)$ are $2\pi$-periodic by definition. This trouble seems to be the main reason why many authors look for $2\pi$-invariant position operators in order to construct relevant uncertainty measures on the circle.

The first such operators used probably were $\sin\varphi$ and $\cos\varphi$ [1]. The variances of these operators satisfy correct inequalities [1]

$$(\Delta p_\varphi)^2 (\Delta \sin\varphi)^2 \geq |\langle \cos\varphi \rangle|^2/4, \qquad (\Delta p_\varphi)^2 (\Delta \sin\varphi)^2 \geq |\langle \cos\varphi \rangle|^2/4 \qquad (3)$$

However one easily find states in which the variances $(\Delta \sin\varphi)^2$ and $(\Delta \cos\varphi)^2$ take values greater than the corresponding ones in the uniform distribution $p_{\text{uni}}(\varphi) = 1/2\pi = |\psi_m(\varphi)|^2$: in $\psi_m(\varphi)$ one has $(\Delta \sin\varphi)^2 = (\Delta \cos\varphi)^2 = 1/2$, while in $\psi_{\cos}(\varphi) = (1/\sqrt{\pi})\cos\varphi$ these variances are $(\Delta \cos\varphi)^2 = 3/4$, $(\Delta \sin\varphi)^2 = 1/4$. In $\psi_{\sin}(\varphi) = (1/\sqrt{\pi})\sin\varphi$ they are interchanged – $(\Delta \cos\varphi)^2 = 1/4$, $(\Delta \sin\varphi)^2 = 3/4$. The two states $\psi_{\cos}(\varphi)$ and $\psi_{\sin}(\varphi)$ coincide under the shift $\varphi \to \varphi \pm \pi/2$, therefore it is reasonable to have coinciding (or close) measures of spread for them, which should be less than those in the eigenstates $\psi_m(\varphi)$. These deficiencies are partially removed by the "uncertainty measure" [1] $(\tilde{\Delta}\varphi)^2 = (\Delta\cos\varphi)^2 + (\Delta\sin\varphi)^2$, which can be written also in the forms

$$(\tilde{\Delta}\varphi)^2 = 1 - \langle \cos\varphi \rangle^2 - \langle \sin\varphi \rangle^2 = 1 - |\langle U(\varphi) \rangle|^2, \qquad U(\varphi) = e^{i\varphi} \qquad (4)$$

The quantity $\tilde{\Delta}\varphi$ has been considered also in [5] and [3]. In [3] it was noted that $\tilde{\Delta}\varphi$ has the meaning of radial distance of the centroid of the ring distribution $p(\varphi)$ from the circle line (and $\langle \cos\varphi \rangle^2 + \langle \sin\varphi \rangle^2$ is the squared centroid's distance from the center of the circle – see figure 1 in [3]). From (4) and (3) it follows that [1]

$$(\Delta p_\varphi)^2 (\tilde{\Delta}\varphi)^2 \geq \frac{1}{4}(\langle \cos\varphi \rangle^2 + \langle \sin\varphi \rangle^2) \qquad (5)$$

This UR is approximately minimized in the canonical coherent states (CS) $|\alpha, \beta\rangle$ of the two dimensional oscillator with large value of $\text{Re}^2\alpha + \text{Re}^2\beta$ [1].

However if one consider the quantity $(\tilde{\Delta}\varphi)^2$, eq. (4), as a delocalization measure on the circle one encounters some unsatisfactory results. For example, it produces the same maximal delocalization (i.e. $\tilde{\Delta}\varphi = 1$) for the eigenstates $\psi_m(\varphi)$ of $\hat{p}_\varphi$ and for all states $\psi(\varphi)$ with the property $|\psi(\varphi)| = |\psi(\varphi + \pi)|$. The centroid for those $\pi$-periodic distributions $|\psi(\varphi)|^2$ is in the center of the ring. On figure 1 graphics of three $\pi$-periodic distributions are shown: uniform one $p_{\text{uni}}(\varphi) = 1/2\pi = |\psi_m(\varphi)|^2$, $p_{\sin}(\varphi) = \psi_{\sin}(\varphi)^2 = \sin^2\varphi/\pi$ and $p_{\sin2}(\varphi) = \sin(2\varphi)^2/\pi$. It is clear that the localization of that distributions is quite different, and it is desirable to have an uncertainty measure that distinguishes between them.

A rather nonstandard expressions for position and angular momentum uncertainties for a particle on the circle were introduced and discussed in [9]:

$$\Delta^2(\hat{p}_\varphi) = \frac{1}{4}\ln(\langle e^{-2\hat{p}_\varphi}\rangle\langle e^{2\hat{p}_\varphi}\rangle), \quad \Delta^2(\hat{\varphi}) = -\frac{1}{4}\ln|\langle U(\varphi)^2\rangle|^2 \qquad (6)$$

For a large sets of states these quantities obey the inequality $\Delta^2(\hat{p}_\varphi) + \Delta^2(\hat{\varphi}) \geq 1$, the equality being reached in the eigenstates $|\xi\rangle$ of the operator $Z = \exp(-\hat{p}_\varphi + 1/2)U(\varphi)$. The family of $|\xi\rangle$ is overcomplete and the states $|\xi\rangle$ are called CS on the circle [8, 5, 9].

The functional $\Delta^2(\hat{\varphi})$, based on the $2\pi$-invariant operator $U(\varphi)$ ($U(\varphi + 2\pi) = U(\varphi)$) was proposed as a position uncertainty on the circle. However this uncertainty measure was found [13] to be not quite consistent with state localization: on CS $|\xi\rangle$ it equals $1/2$, while on the visually worse localized states $|\xi\rangle - |-\xi\rangle$ (Schrödinger cat states on the circle) it can take rather less value of $0.33$ (see [13] and figure 2 therein). On the above noted states $\psi_{\sin}(\varphi)$, $\psi_{\sin2}(\varphi)$ and $\psi_m(\varphi)$ it takes values $0.346, \infty, \infty$. Thus it makes distinction between $\psi_{\sin}(\varphi)$ and $\psi_{\sin2}(\varphi)$ and $\psi_m(\varphi)$, but identifies $\psi_{\sin2}(\varphi)$ with the uniform state $\psi_m(\varphi)$ (see figure 1).

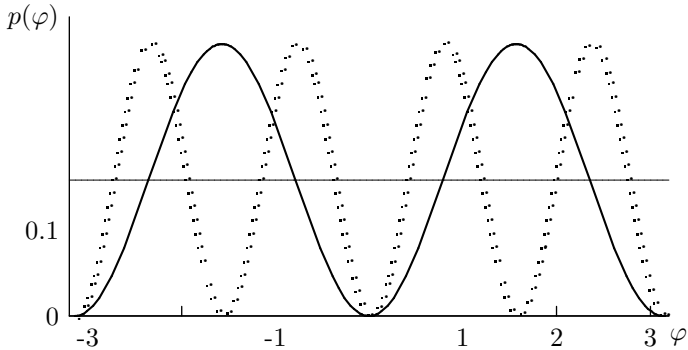

**Figure 1.** $\pi$-periodic, $\pi/2$-periodic and uniform distributions on the circle: $p_{\sin}(\varphi) = (\sin\varphi)^2/\pi$ (dot line), $p_{\sin2}(\varphi) = (\sin(2\varphi))^2/\pi$ (solid line) and $p_{\text{uni}}(\varphi) = 1/2\pi = |\psi_m(\varphi)|^2$. The functional $\tilde{\Delta}\varphi$, eq. (4), on all these distributions takes the same maximal value of 1, while $\Delta^2(\hat{\varphi})$, eq. (6), takes the values $0.346, \infty$ and $\infty$ respectively.

New position uncertainty measures, that are better consistent with the localization on the circle are constructed in the next section as particular cases ($\vartheta = \pi/2$) of the measures on the sphere.

## 3. Uncertainty measures on the sphere

A point on the sphere $\mathbb{S}^2$ is determined by the spherical angles $\varphi$ and $\vartheta$. The Hilbert space of states for a particle on the sphere is defined as the space of square integrable functions on $\mathbb{S}^2$ with respect to the normalized measure (on the unit sphere) $d\mu(\varphi, \vartheta) = \sin\vartheta \, d\vartheta \, d\varphi/4\pi \equiv dS/4\pi$. Wave functions $\psi(\vartheta, \varphi)$ have to be $2\pi$-periodic in $\varphi$ (periodicity in $\vartheta$ is not required).

The measure of uncertainty of $\vartheta$ in a state $\psi(\vartheta, \varphi)$ may be adopted as the ordinary variance $(\Delta\vartheta)^2 = \langle\psi|\vartheta^2|\psi\rangle - \langle\psi|\vartheta|\psi\rangle^2$. The uncertainty measure for $\varphi$ can not be taken in a similar way. In view of the nonperiodicity of $\varphi$ the standard "means" $\langle\varphi^k\rangle$, $k = 1,\ldots$, are ill defined: their values depend on the limit $\varphi_0$ of integration (on the unit sphere: $dS = \sin\vartheta d\vartheta d\varphi$,

$$\langle\psi|\varphi^k|\psi\rangle = \int_0^\pi \sin\vartheta \mathrm{d}\vartheta \int_{\varphi_0-\pi}^{\varphi_0+\pi} \varphi^k|\psi(\varphi,\vartheta)|^2 \mathrm{d}\varphi \equiv M^{(k)}\varphi(\varphi_0) \qquad (7)$$

Following the scheme of refs [13] for the case of a circle we define the $\varphi$-uncertainty measure on the sphere as

$$(_c\Delta\varphi)^2 = \int_0^\pi \sin\vartheta \mathrm{d}\vartheta \int_{\varphi_c-\pi}^{\varphi_c+\pi} \varphi^2|\psi(\varphi,\vartheta)|^2 \mathrm{d}\varphi - \left(\int_0^\pi \sin\vartheta \mathrm{d}\vartheta \int_{\varphi_c-\pi}^{\varphi_c+\pi} \varphi|\psi(\varphi,\vartheta)|^2 \mathrm{d}\varphi\right)^2 \qquad (8)$$

where $\varphi_c$ is the $\varphi$-coordinate of the *center of the packet* $|\psi(\varphi,\vartheta)|^2$. For packets $|\psi(\varphi,\vartheta)|$ that *are not* $2\pi/k$-periodic, $k = 2, \ldots$, in $\varphi$ the angle $\varphi_c$ can be determined as the polar angle of a pont in the plane with cartesian coordinates $x_c = \langle\cos\varphi\rangle$, $y_c = \langle\sin\varphi\rangle$. For packets that are $2\pi/k$-periodic, $k = 2, \ldots$, in $\varphi$, one obtains $x_c = 0 = y_c$, i.e. the center of the packet $\varphi_c$ remains undefined in this way. To overcome this difficulty suffice it to observe that the packet center $\varphi_c$, when determined from the above $x_c$ and $y_c$, satisfies the conditions (as checked on several examples)

$$M\varphi(\varphi_c) = \varphi_c,$$
$$\int_0^\pi \sin\vartheta|\psi(\varphi_c + \pi, \vartheta)|^2 \mathrm{d}\vartheta \leq \frac{1}{2\pi} \qquad (9)$$

where $M\varphi(\varphi_0)$ is the limit-dependent "mean" of $\varphi$, defined in eq. (7) for $k = 1$. Therefore it is reasonable to define $\varphi$-coordinate $\varphi_c$ of the center of the wave packet on the sphere as solution of the system (9). It is straightforward to check that conditions (9) ensure *the minimum* of the limit-dependent variance $\Delta\varphi(\varphi_0)$ as a function of $\varphi_0$,

$$(\Delta\varphi)^2(\varphi_0) = \int_0^\pi \sin\vartheta \mathrm{d}\vartheta \int_{\varphi_0-\pi}^{\varphi_0+\pi} \varphi^2|\psi(\varphi,\vartheta)|^2 \mathrm{d}\varphi - (M\varphi(\varphi_0))^2 \qquad (10)$$

One can verify that the function $(\Delta\varphi)^2(\varphi_0)$ is $2\pi$-periodic [i.e. $(\Delta\varphi)^2(\varphi_0 + 2\pi) = (\Delta\varphi)^2(\varphi_0)$], therefore the minimum always exists.

For $2\pi/k$-periodic, $k = 2, \ldots$, in $\varphi$ wave packets the conditions (9) may have more than one solutions (in fact $k$ solutions). We call these $2\pi/k$-periodic packets *multi-centered*.

The $\vartheta$-coordinate $\vartheta_c$ of the packet center on the sphere can be defined as the mean of $\theta$:

$$\vartheta_c = \int \vartheta|\psi(\varphi,\vartheta)|^2 \mathrm{d}S.$$

Thus the coordinates of the wave packet center on the sphere are $(\varphi_c, \vartheta_c)$. The uncertainty (or delocalization) measure of a state $|\psi\rangle$ on the sphere can be defined in two complementary ways: as a sum or as a product of the corresponding $\varphi$- and $\vartheta$- measures,

$$M_+(\psi) := (_c\Delta_\psi\varphi)^2 + (\Delta_\psi\vartheta)^2, \quad \text{or} \quad M_\bullet(\psi) := (_c\Delta_\psi\varphi)^2(\Delta_\psi\vartheta)^2 \qquad (11)$$

**Examples.** Let us illustrate the relevance of the above constructed position uncertainty measures on two families of states $f_{uv\gamma k}(\varphi,\vartheta)$ and $\psi_{uv\tau}(\varphi,\vartheta)$,

$$f_{uv\gamma k}(\varphi,\vartheta) = N(u,v,\gamma,k)\left[2 + \cos(k\varphi - u) + \cos(3(\vartheta - v)/2)\right]^\gamma \qquad (12)$$

$$\psi_{uv\tau}(\varphi,\vartheta) = N(u,v,\tau)\sum_{l=0}^\infty e^{-\tau l(l+1)/2}\sqrt{2l+1}\,P_l(\cos\theta) \qquad (13)$$

where $N(u,v,\gamma,k), N(u,v,\tau)$ are normalization factors, $\gamma$ and $\tau$ are real positive parameters, $k$ is a positive integer, $P_l(x)$ are Legendre polynomials, and $\theta$ is the angle between radii of the current point $(\varphi,\vartheta)$ and a fixed point $(u,v)$ on the unit sphere.

The function $f_{uv\gamma k}(\varphi, \vartheta)$ is constructed as a $k$-peak state on the unit sphere ($k = 1, 2, \ldots$), the width of the peaks being decreasing with $\gamma$. Thus $1/\gamma$ plays a role of a delocalization parameter, and $\gamma$ – a *squeezing parameter* of the states $f_{uv\gamma k}$ (see figures 2 and 3 for the cases of $k = 2$, $u = \pi, v = \pi/2$, and $\gamma = 1$ (figure 2) and $\gamma = 5$ (figure 3) (or $1/\gamma = 1$ and $0.2$)). In these states the above defined position uncertainty measures are calculated as $({}_c\Delta\varphi)^2 = 2.94$, $(\Delta\vartheta)^2 = 0.329$ and $({}_c\Delta\varphi)^2 = 2.57$, $(\Delta\vartheta)^2 = 0.146$ correspondingly. Both ${}_c\Delta\varphi$ and $\Delta\vartheta$ (and $M_+$ and $M_\bullet$ as well) are found decreasing when $1/\gamma \to 0$, i.e. $\gamma$ indeed appears as a *squeezing parameter* (the greater is $\gamma$ the stronger is squeezing of the position uncertainty measures). The packet centers $(\varphi_c, \vartheta_c)$ do not depend on $\gamma$ and for $u = \pi, v = \pi/2, k = 2$ the two centers are $(0, \pi/2)$ and $(\pi, \pi/2)$.

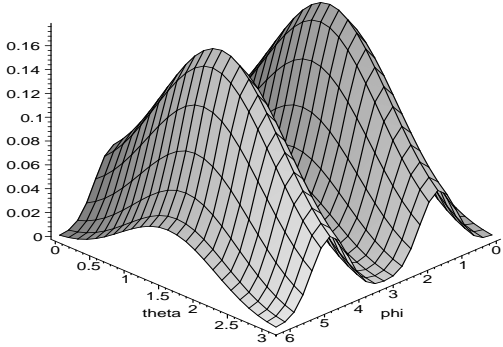
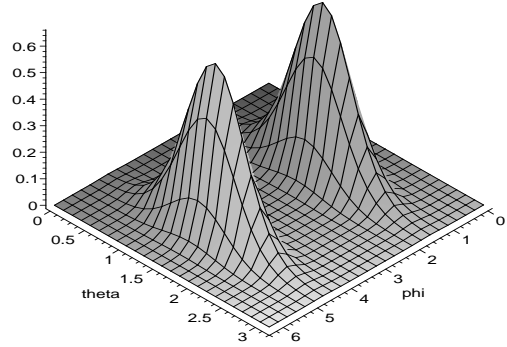

**Figure 2.** Plot of the two peak distribution $|f_{uv\gamma k}(\varphi, \vartheta)|^2$ with $u = \pi, v = \pi/2$, $k = 2$, $\gamma = 1$. In this state the $\varphi$- and $\vartheta$-uncertainties are $({}_c\Delta\varphi)^2 = 2.94$, $(\Delta\vartheta)^2 = 0.329$. The two packet centers are at $(\varphi = 0, \vartheta = \pi/2)$ and $(\varphi = \pi, \vartheta = \pi/2)$.

**Figure 3.** Plot of the two peak distribution $|f_{uv\gamma k}(\varphi, \vartheta)|^2$ with $u = \pi, v = \pi/2$, $k = 2$, $\gamma = 5$. In this state the $\varphi$- and $\vartheta$-uncertainties are $({}_c\Delta\varphi)^2 = 2.57$, $(\Delta\vartheta)^2 = 0.146$. The two packet centers are at $(\varphi = 0, \vartheta = \pi/2)$ and $(\varphi = \pi, \vartheta = \pi/2)$.

The function $\psi_{uv\tau}$ is taken from paper [6], where the set $\{\psi_{uv\tau}\}$ is shown to form an overcomplete set of states (for every $\tau$; and $u, v$ may be complex) on the sphere $\mathbb{S}^2$, called *coherent states on the sphere* (CS on the sphere). In [6] CS are constructed on $d$-dimensional sphere, and CS on $\mathbb{S}^2$ were previously constructed in [10]. The shapes of CS $\psi_{uv\tau}$ are shown on figures 4 and 5 for the cases of $u = \pi, v = \pi/2$, and $\tau = 1$ (figure 4) and $\tau = 0.2$ (figure 5). Calculations show that the less $\tau$ is the less is the area of $\mathbb{S}^2$ in which CS are concentrated, confirming the suggestion of [6]. In these CS the above defined position uncertainty measures take the values $({}_c\Delta\varphi)^2 = 1.57$, $(\Delta\vartheta)^2 = 0.419$ (for $\tau = 1$) and $({}_c\Delta\varphi)^2 = 0.439$, $(\Delta\vartheta)^2 = 0.185$ (for $\tau = 0.2$). Both ${}_c\Delta\varphi$ and $\Delta\vartheta$ are decreasing (thereby $M_+$ and $M_\bullet$ also are decreasing) when $\tau \to 0$, i.e. $1/\tau$ appears as a position squeezing parameter. The packet center $(\varphi_c, \vartheta_c)$ does not depend on $\tau$, and for CS with $u = \pi, v = \pi/2$ it is $(\pi, \pi/2)$.

It is worth noting that the shapes of one-peak states $f_{uv\gamma 1}(\varphi, \vartheta) \equiv f_{uv\gamma}(\varphi, \vartheta)$ with $\gamma = 1/\tau$ and CS $\psi_{uv\tau}(\varphi, \vartheta)$ are quite similar: the packet centers of both states are determined by $u, v$, and the position uncertainties ${}_c\Delta\varphi$ and $\Delta\vartheta$ vary with $\tau$ similarly. In particular in $f_{uv\gamma}$ with $u = \pi, v = \pi/2$ and $1/\gamma = \tau = 1, 0.2$ one finds $({}_c\Delta\varphi)^2 = 1.91, 0.418$, $(\Delta\vartheta)^2 = 0.329, 0.146$ respectively, which are to be compared with $({}_c\Delta\varphi)^2 = 1.57, 0.439$, $(\Delta\vartheta)^2 = 0.419, 0.185$ in the corresponding CS $\psi_{uv\tau}$.

Due to the factor $\sin\vartheta$ in the surface element $dS$ the most delocalized states on the sphere is $\psi_\alpha(\varphi, \vartheta) = \exp(i\alpha(\varphi, \vartheta))/(\pi\sqrt{2\sin\vartheta})$ (and not the uniform one $\psi_{\rm uni} = 1/\sqrt{4\pi}$): In $\psi_\alpha$ the $\varphi$- and $\vartheta$-uncertainty measures take the values $({}_c\Delta\varphi)^2 = (\Delta\varphi)^2 = \pi^2/3 \simeq 3.29$, $(\Delta\vartheta)^2 = \pi^2/12 \simeq 0.82$, while in $\psi_{\rm uni}$ $({}_c\Delta\varphi)^2 = \pi^2/3$, $(\Delta\vartheta)^2 = \pi^2/4 - 0.2 \simeq 0.47$.

As we have already noted position uncertainty measures are positive maps of states (in fact of the corresponding probability distributions), associated with coordinate variables. It is then clear that one can construct such measures using other coordinates $x_i(\varphi, \vartheta), i = 1, 2$, on the sphere, such as the stereographic projections $q_i$ and the "wrapping" coordinates $(\eta, \xi)$ [4].

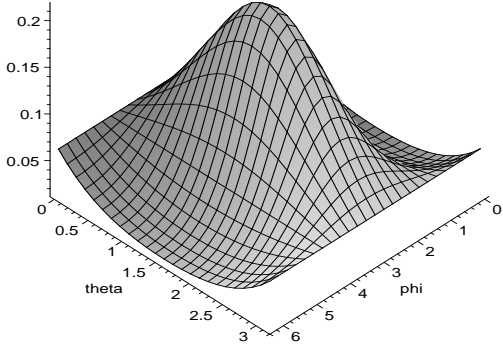 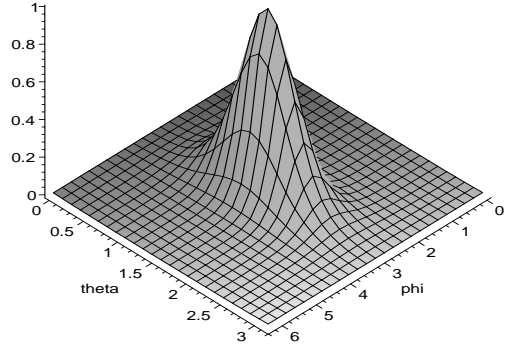

**Figure 4.** Plot of CS distribution on the sphere $|\psi_{uv\tau}(\varphi,\vartheta)|^2$ with $u=\pi, v=\pi/2, \tau=1$. In this state the $\varphi$- and $\vartheta$-uncertainties are $(_c\Delta\varphi)^2 = 1.57$, $(\Delta\vartheta)^2 = 0.419$. The packet center is at $(\varphi=\pi, \vartheta=\pi/2)$.

**Figure 5.** Plot of CS distribution on the sphere $|\psi_{uv\tau}(\varphi,\vartheta)|^2$ with $u=\pi, v=\pi/2, \tau=0.2$. In this state the $\varphi$- and $\vartheta$-uncertainties are $(_c\Delta\varphi)^2 = 0.439$, $(\Delta\vartheta)^2 = 0.185$. The packet center is at $(\varphi=\pi, \vartheta=\pi/2)$.

However when $x_i(\varphi,\vartheta)$ depend on $\varphi$ through $\cos\varphi$ and/or $\sin\varphi$ (the case of $q_i$ and $(\eta,\xi)$), the standard variance will exhibit deficiencies like those on the circle, discussed in previous section. The stereographic coordinates $q_i$,

$$q_1 = 2r\cot(\vartheta/2)\cos\varphi, \quad q_2 = 2r\cot(\vartheta/2)\sin\varphi \tag{14}$$

exhibit an extra deficiency (coming from the factor $\cot(\vartheta/2)$) – their variances are infinite in all states $\psi(\varphi,\vartheta)$ which do not vanish sufficiently fast when $\vartheta \to 0$. Examples of such states are $f_{uv\gamma k}(\varphi,\vartheta)$ and CS $\psi_{uv\tau}(\varphi,\vartheta)$ discussed above. These states do not vanish at $\vartheta = 0$. Therefore $\langle q_i^2 \rangle = \infty$. In the most delocalized states $\psi_\alpha(\vartheta)$ the averages of $q_i^2$ are also divergent.

## 4. Uncertainty inequalities on the sphere

The uncertainty relations (UR's) for the position delocalization measure on the sphere encounter two problems. The first one is related to a ill property of the multiplication position operator $\varphi$: the function $\varphi\psi(\varphi,\vartheta)$ is no more $\pi$-periodic. The second subtle comes from the non-hermiticity of the operator $-i\partial/\partial\vartheta$ ($\hbar=1$). There is a third problem (compared to the case of a circle) related to several position and momentums operators for a particle on the sphere: one has to formulate uncertainty relations for several observables.

In order to overcome these difficulties we have to apply the scheme of the Gram-Robertson matrix, developed in [14]. For one state $|\psi\rangle$ and several observables $X_i$, $i=1,\ldots,n$, the Gram-Robertson matrix $G$ is defined as [14] $G = \{G_{ij}\}$,

$$G_{ij} = \langle (X_i - \langle X_i\rangle)\psi | (X_j - \langle X_j\rangle)\psi\rangle \equiv G_{ij}(\psi; \vec{X}) \tag{15}$$

It was shown that the characteristic coefficients of the symmetric part $S$ of $G$ ($S = (G+G^T)/2$) are greater or equal to that of the antisymmetric part $A$ ($A = -i(G-G^T)/2$, $G^T$ being the transposed $G$). These inequalities are called (generalized) *characteristic* UR's for the $n$ observables $X_i$ in a state $|\psi\rangle$ [14]. In particular, the senior characteristic UR reads

$$\det S \geq \det A. \tag{16}$$

The real and symmetric matrix $S$ is defined as a (generalized) uncertainty matrix, and $A$ is regarded as a generalization of the matrix of mean commutators $-i\langle[X_j, X_k]\rangle$. When the action of $X_i$ is well defined on $X_j|\psi\rangle$ the matrix $S$ coincide with the standard uncertainty (or covariance) matrix $\{\langle\psi|(X_i - \langle X_i\rangle)(X_j - \langle X_j\rangle)\psi\rangle\} = \{\text{Cov}(X_i, X_j)\} \equiv \sigma_{ij})$, where $\text{Cov}(X_i, X_j)$ is the standard covariance of $X_i$ and $X_j$. The other notations for the covariance $\text{Cov}(X_i, X_j)$ and variance $\text{Cov}(X_i, X_i)$ are $\Delta X_i X_j$ and $(\Delta X_i)^2$. In such "smooth" cases the senior characteristic inequality (16) reads

$$\det \sigma \geq \det C, \quad \text{where} \quad C = \{-\mathrm{i}\langle[X_k, X_j]\rangle/2\} \tag{17}$$

and this latter inequality was first obtained by Robertson [11]. It is a generalization of the Schrödinger (or Schrödinger-Robertson) UR for two observable $X_1$, $X_2$ (first established in [12]),

$$(\Delta X_1)^2(\Delta X_2)^2 \geq \frac{1}{4}|\langle[X_1, X_2]\rangle|^2 + (\text{Cov}(X_1, X_2))^2 \tag{18}$$

If for some reason the repeated action $X_i X_j|\psi\rangle$ is not correctly defined one has to resort to generalized UR (16), which for two operators reads

$$(_g\Delta X_1)^2(_g\Delta X_2)^2 \geq \frac{1}{4}|_g\langle[X_1, X_2]\rangle|^2 + (_g\text{Cov}(X_1, X_2))^2 \tag{19}$$

where $_g\text{Cov}(X_1, X_2) = \text{Re}\langle(X_1-\langle X_1\rangle)\psi|(X_2-\langle X_2\rangle)\psi\rangle$, $(_g\Delta X)^2 = {_g}\text{Cov}(X, X)$, and $_g\langle[X_1, X_2]\rangle = -2i\text{Im}\langle(X_1 - \langle X_1\rangle)\psi|(X_2 - \langle X_2\rangle)\psi\rangle$ [14]. The UR (19) quite similar to (18). Thus we may define the generalized covariance and the generalized mean commutator as $_g\text{Cov}(X_1, X_2)$ and $_g\langle[X_1, X_2]\rangle$ respectively [14, 2].

Let us note that for two operators the inequality $\det S \geq \det A$, eq. (16), is equivalent to $\det G \geq 0$, and UR (18) is equivalent to $\det(\sigma + \mathrm{i}C) \geq 0$.

The two Hermitian operators on the sphere $X_1 = \varphi$ and $X_2 = -\mathrm{i}\partial/\partial\varphi \equiv \hat{p}_\varphi$ constitute an example in which of $X_2 X_1 \psi(\varphi, \vartheta)$ is not properly defined: $\hat{p}_\varphi$ is not Hermitian on states $\varphi\psi(\varphi, \vartheta)$, since $\psi' = \varphi\psi(\varphi, \vartheta)$ is not $2\pi$-periodic. Therefore the $\varphi$–$p_\varphi$ UR should resort to eq. (19). However even the generalized covariance of $\varphi$, $\hat{p}_\varphi$ and the generalized variance of $\varphi$ depend on the limits of integration when calculating means like

$$\langle\varphi\rangle = \int_{\varphi_0-\pi}^{\varphi_0+\pi} \varphi|\psi|^2 \mathrm{d}S.$$

Fortunately, the variance of $\varphi$ is $2\pi$-periodic function of $\varphi_0$, therefore its global extrema exist and we may define $\varphi$-uncertainty measure on the sphere $_c\Delta\varphi$ as explained in the previous section. Then the $\varphi$–$p_\varphi$ UR on the sphere could be adopted in the form

$$(_c\Delta\varphi)^2 (\Delta p_\varphi)^2 \geq |_c\langle(\varphi - \langle\varphi\rangle)\psi|(\hat{p}_\varphi - \langle\hat{p}_\varphi\rangle)\psi\rangle|^2 \tag{20}$$

where $_c\langle X\rangle$ means that the average of $X$ is calculated by integration from $\varphi_c - \pi$ to $\varphi_c + \pi$, $\varphi_c$ being the wave packet center. Note that the right hand side of (20) may vanish (on the eigenstates of $p_\varphi$ for example), so that the less precise version of the inequality (20) is $(_c\Delta\varphi)^2 (\Delta p_\varphi)^2 \geq 0$. The variance $\Delta p_\varphi$ should be called *complementary measure* to the position delocalization measure $_c\Delta\varphi$.

A natural definition of a *complementary measure* is the following: a state measure $M(\psi)$ is a complementary one to a state measure $N(\psi)$ if $M(\psi)$ tends to its global maximum (minimum) when $N(\psi)$ tends to its global minimum (maximum). The variance of $\hat{p}_\varphi = -\mathrm{i}\partial/\partial\varphi$ is a complementary one to the delocalization measure $(_c\Delta\varphi)^2$. For a particle on the real line the variances of the coordinate $x$ and momentum $\hat{p} = -\mathrm{i}\mathrm{d}/\mathrm{d}x$ are complementary measures. It is clear that a given measure $N(\psi)$ may have many complementary measures. Additional criteria should be used to specify the most convenient complementary measure is every special case. Let us also note, that measures map states on the positive part on the real line, i.e. these are many-to-one maps. Therefore they may reach their extremal values on a large subset of states.

Our aim now is to construct measure complementary to the well defined $\vartheta$-uncertainty measure (the variance) $(\Delta\vartheta)^2$ on the sphere. The problem with such complementary measure, and the $\vartheta$–$p_\vartheta$ UR as well, is in the ill property of the operator $-\mathrm{i}\partial/\partial\vartheta \equiv \tilde{p}_\vartheta$: this operator obey formally the relation $[\vartheta, \tilde{p}_\vartheta] = \mathrm{i}$, however it is not Hermitian. Therefore the variance $\langle\tilde{p}_\vartheta^2\rangle - \langle\tilde{p}_\vartheta\rangle^2$ may be complex.

It is easy to point out Hermitian operator $\hat{p}_{0\vartheta}$ with the same commutator as for $\vartheta$ and $\tilde{p}_\vartheta$. Such is the operator $\hat{p}_{0\vartheta} = -\mathrm{i}\partial/\partial\vartheta - (\mathrm{i}/2)\cot(\vartheta)$. One has $[\varphi, \hat{p}_{0\vartheta}] = \mathrm{i}$. Then we can write the standard Schrödinger UR $(\Delta\vartheta)^2 (\Delta p_{0\vartheta})^2 \geq 1/4 + (\mathrm{Cov}(p_{0\vartheta}, \vartheta))^2$.

However the variance $(\Delta\hat{p}_{0\vartheta})^2$ could hardly serve as a *complementary measure* to the position measure $(\Delta\vartheta)^2$ since it is diverged in all states that are not vanishing at $\vartheta = 0$, and $\vartheta = \pi$. Examples of such states are $f_{uv\gamma k}(\varphi, \vartheta)$ and CS $\psi_{uv\tau}(\varphi, \vartheta)$, treated in section 3. This ill property of $\hat{p}_{0\vartheta}$ stems from the fact that the functions $\hat{p}_{0\vartheta}\psi(\varphi, \vartheta)$ are not normalizable (where $\psi(\varphi, \vartheta)$ represent normalized state).

Fortunately, there are simple Hermitian operators the variances of which could be regarded as complementary to $(\Delta\vartheta)^2$ measures. These are the first order differential operators $\hat{p}_{n\vartheta}$ of the form

$$\hat{p}_{n\vartheta} = -\mathrm{i}\sin^n\vartheta\frac{\partial}{\partial\vartheta} - \mathrm{i}\frac{n+1}{2}\cos\vartheta\sin^{n-1}\vartheta, \quad n = 1, 2, \ldots \tag{21}$$

We have $[\vartheta, \hat{p}_{n\vartheta}] = \mathrm{i}\sin^n\vartheta$, and the transformed states $\hat{p}_{n\vartheta}\psi(\varphi, \vartheta)$ are normalizable for all $n \geq 1$. The variances and covariances of $\vartheta$ and $\hat{p}_{n\vartheta}$ satisfy the Schrödinger UR (18). In view of $\sin\vartheta \geq 0$ in the interval $(0, \pi)$ we have $\langle\sin^n\vartheta\rangle > 0$, therefore the right hand side of Schrödinger UR never vanishes, i.e. $(\Delta\vartheta)^2(\Delta p_{n\vartheta})^2 > 0$. This is a proof that the spectrum of operators $\hat{p}_{n\vartheta}$ is not discrete. Finally we have to point out which (from all $(\Delta p_{n\vartheta})^2$) is the *best* complementary measure to the position uncertainty measure $(\Delta\vartheta)^2$. We have to apply some criterions. One such criterion could be the demand that the complementary measure $(\Delta p_{n\vartheta})^2$ be minimal (with respect to $n$) in the most delocalized state $\psi_{\alpha=0}(\varphi, \vartheta) = 1/(\pi\sqrt{2\sin\vartheta}) \equiv \psi_0(\vartheta)$. Numerical calculations show that this criterion selects $\Delta p_{1\vartheta}$ and $\Delta p_{2\vartheta}$:

In $\psi_0(\vartheta)$ we find $(\Delta p_{n>2,\vartheta})^2 > (\Delta p_{2\vartheta})^2 \simeq (\Delta p_{1\vartheta})^2 \simeq 0.125$. Note that the variances of $\hat{p}_{n\vartheta}$ on $\psi_\alpha(\varphi, \vartheta)$ do depend on the phase $\alpha$ when the latter is a function of angle $\vartheta$.

Another natural criterion is the lower limit ot the product $(\Delta p_{n\vartheta})^2(\Delta\vartheta)^2$, for which one has the standard UR

$$(\Delta p_{n\vartheta})^2(\Delta\vartheta)^2 \geq |\langle\sin^n\vartheta\rangle|^2/4 \tag{22}$$

It is clear that in any state the inequality $|\langle\sin^n\vartheta\rangle| > |\langle\sin\vartheta\rangle|$ holds. So the second criteria picks up from the set $\{(\Delta p_{n\vartheta})^2 : n = 1, 2, \ldots\}$ the variance $(\Delta p_{1\vartheta})^2$ as the *best complementary measure* to the position delocalization measure $(\Delta\vartheta)^2$. In analogy with $(\Delta\vartheta)^2$ we may denote this $\vartheta$-complementary measure as $(\Delta p_\vartheta)^2$, i.e. we put $(\Delta p_{1\vartheta})^2 \equiv (\Delta p_\vartheta)^2$. In the two states $f_{uv\gamma k}(\varphi, \vartheta)$ and two CS $\psi_{uv\tau}(\varphi, \vartheta)$ represented in figures 2, 3 and 4, 5 the values of $(\Delta p_{1\vartheta})^2$ are $0.57, 1.54$ and $0.419, 1.38$ respectively. It is remarkable that in CS $\psi_{\pi\pi/2\tau}$ with $\tau = 1$ the $\vartheta$-position uncertainty is approximately equal to the complementary one: $(\Delta\vartheta)^2 = 0.419 = (\Delta p_{1\vartheta})^2$. This is to be compared with the case of CS on the plane, where the two complementary uncertainties (position and momentum uncertainties) are equal (in any CS however).

Thus the Hermitian operator $\hat{p}_{1\vartheta}$ could be examined as a momentum $\hat{p}_\vartheta$ complementary to the variable $\vartheta$. The four measures on the sphere $(_c\Delta\varphi)^2$, $(\Delta\vartheta)^2$, $(\Delta p_\varphi)^2$, and $(\Delta p_\vartheta)^2$ satisfy the generalized Robertson UR (16), where the integration with respect to $\vartheta$ in all matrix elements involving the variable $\varphi$ should be from $\varphi_c - \pi$ to $\varphi_c + \pi$.

Let us note that the means $\langle\psi_0|\hat{p}_i|\psi_0\rangle$, where $\hat{p}_i$ are the two Hermitian operators [4], "conjugated" to the stereographic coordinates $q_i$ (14), are divergent. Thus these operators move the most delocalized state $\psi_0(\vartheta)$ (and many other states as well) away from the appropriate Hilbert space, and their variances are not convenient as complementary to the position uncertainty measures on the sphere.

## 5. Conclusion

We have constructed two position uncertainty measures $(_c\Delta\varphi)$, $(\Delta\vartheta)^2$ and two related complementary measures $(\Delta p_\varphi)^2$, $(\Delta p_\vartheta)^2$ for a particle on the sphere. The $\vartheta$-complementary measure $(\Delta p_\vartheta)^2$ is a variance of the new operator $\hat{p}_{1\vartheta}$, eq.(21).

The four measures obey the genralized Robertson UR (16), any two of them satisfying the Schrödinger-Robertson type UR (19). The relevance of the constructed measures are illustrated on the example of two sets of states: $f_{uv\gamma k}(\varphi, \vartheta)$, eq. (12), and coherent states on the sphere [6] $\psi_{uv\tau}(\varphi, \vartheta)$, eq. (13). The relevance of $(_c\Delta\varphi)^2$ as a position measure should not be considered as a proof that the right position operator $\hat{\varphi}$ for the azimuthal angle is the multiplication by $\varphi$.

The presented approach to uncertainty measures on $\mathbb{S}^2$ could be easily extended to higher dimensional spheres.